\documentclass[aps,prl,twocolumn,nofootinbib,groupedaddress]{revtex4-2}

\usepackage[english]{babel}
\usepackage{graphicx}
\usepackage{amsmath,amssymb,amsbsy,amstext,amsfonts}
\usepackage{bm}
\usepackage{xcolor}
\usepackage{hyperref}
\hypersetup{colorlinks=true,linkcolor=purple,filecolor=purple,urlcolor=purple,citecolor=purple}

\renewcommand{\d}{\mathrm{d}}
\renewcommand{\Re}{\mathrm{Re}}

\newcommand{\F}{\mathcal{F}}

\begin{document}

\title{Laplace Space for Cosmological Correlators}

\author{Nathan Belrhali}
\author{Arthur Poisson}
\author{S\'ebastien Renaux-Petel}
\affiliation{Institut d'Astrophysique de Paris, CNRS, Sorbonne Universit\'e,
98 bis bd Arago, 75014 Paris, France}

\begin{abstract}
Deep inside the horizon, every cosmological mode oscillates as a flat-space plane wave. A Laplace transform turns this fact into a general method: it resolves each curved-space mode into a superposition of plane waves dressed by a kernel that encodes the spacetime geometry, field content and dynamics, collapsing the time integrals onto flat-space ones. This provides simple diagrammatic rules that turn cosmological correlator diagrams into their flat-space counterparts integrated against Laplace-space kernels. On the paradigmatic massive single exchange, this integral representation makes the energy singularities manifest and evaluates in closed form as a single, rapidly convergent series valid throughout the kinematic domain, with no patching of separate expansions. The Laplace approach sheds conceptual and computational light on cosmological correlators in virtually any theory of the early universe.

\end{abstract}

\maketitle

\paragraph*{\bf Introduction.}
Cosmological correlators, the late-time $n$-point functions of the primordial curvature and tensor perturbations, are the observables of inflation, holding the promise to uncover primordial physics at energies far beyond the reach of any other probe \cite{Achucarro:2022qrl}. Computing them is demanding: the expansion of the universe breaks time-translation invariance, distorting the mode functions from plane waves into special functions, Hankel functions for a massive field, and leaving the correlators as nested time integrals over products of these. A broad arsenal of techniques has grown around this problem, from the cosmological bootstrap to Mellin-space, spectral, and differential-equation methods~\cite{Arkani-Hamed:2018kmz,Baumann:2019oyu,Baumann:2020dch,Pajer:2020wxk,Jazayeri:2022kjy,Sleight:2019hfp,Sleight:2019mgd,Qin:2022fbv,Goodhew:2020hob,Jazayeri:2021fvk,Melville:2021lst,Hogervorst:2021uvp,Loparco:2023rug,Xianyu:2022jwk,Xianyu:2023ytd,Liu:2024str,Arkani-Hamed:2023kig,Arkani-Hamed:2023bsv,Melville:2024ove,Werth:2024mjg,Arundine:2026fbr,Belrhali:2026ktb,Belrhali:2026rkn,Werth:2023pfl,Pinol:2023oux}.

We develop a different route, rooted in the equivalence principle: locally, any spacetime looks like flat. This geometrical statement takes a sharp dynamical form, as a mode deep inside the Hubble radius does not yet feel the curvature of spacetime and oscillates as a flat-space plane wave, the geometry making itself felt only as the mode is stretched towards the horizon. A Laplace transform turns this observation into a computational tool, resolving a curved-space mode function into a continuous superposition of plane waves and relating it to its elementary flat-space counterpart through a kernel that encodes the spacetime geometry together with the field content and dynamics of the system. In the dual Laplace space the in-in time integrals collapse: each massive line becomes a plane wave and every bulk integral reduces to an elementary flat-space one. Nothing in this mechanism is tied to a particular theory: the same kernel structure dresses essentially any mode function, bringing the whole perturbative machinery of primordial cosmology onto a common flat-space footing.

Two results follow. First, the collapse of the time integrals yields simple diagrammatic rules that produce the Laplace-space integrand directly from a diagram. Second, on the paradigmatic massive single-exchange correlator the method displays a twofold virtue: its integral representation in Laplace space makes the total- and partial-energy singularities manifest ``from flat space'', while it also delivers a single closed-form, very rapidly convergent series valid throughout the entire kinematic domain, with no patching of separate expansions. Such a uniform and fast representation is well suited to scanning shapes of non-Gaussianity, theories, and parameters against cosmological data.

Several groups have shown how cosmological correlators can be obtained by dressing flat-space amplitudes~\cite{Chowdhury:2023arc,Chowdhury:2025ohm,Chowdhury:2026upp,Das:2025qsh,Das:2026vfv,Ansari:2026xkm,Ansari:2026sjf}, but these constructions have so far been confined to massless or conformally coupled fields, setting aside the singularly more involved exchange of massive fields. 
By treating this case, and essentially any theory of interest in primordial cosmology, the Laplace approach opens a new route from flat space to the correlators of the early universe. In this \textit{Letter}, we present the method, derive the diagrammatic rules, carry the single exchange through to closed form, and outline its generality; full derivations and extensions are detailed in a companion paper~\cite{Belrhali:2026laplace}.

\paragraph*{\bf Curved spacetime dynamics from plane waves.}
The perturbative evaluation of cosmological correlators produces nested conformal-time integrals over products of mode functions \cite{Weinberg_2005,Chen:2017ryl}. The elementary object underlying them is a single time integral of a rescaled mode function $\F$ against a plane wave,
\begin{equation}
\hat{\F}(\lambda)=\int_{-\infty(1+i\epsilon)}^{0}\d z\; e^{-i\lambda z}\,\F(z)\,,
\end{equation}
with $z=k_I\tau$, $\lambda=k_E/k_I$ the ratio of the external to the internal momentum. The $i\epsilon$ tilt prepares the Bunch-Davies vacuum that selects the positive-frequency mode functions deep inside the horizon, i.e. $\F(z) \underset{z \to -\infty}{\sim} e^{-iz}$.

Rotating the integration contour onto the imaginary axis identifies $\hat{\F}$ as a Laplace transform, $\hat{\F}(\lambda)=i\,\mathcal{L}[\F(-iz)](\lambda)$, whose two asymptotics fix its analytic data: the late-time, small-$z$ behaviour sets the decay as $\lambda\to\infty$, while the early-time, Bunch-Davies oscillation places a branch point at $\lambda=-1$. Thus $\hat{\F}(\lambda)$ is analytic for $\Re\lambda>-1$ and its analytical continuation carries a single cut along $\lambda\in(-\infty,-1)$. The transform inverts to a Bromwich integral, $i\,\F(-iz)=\int_{c-i\infty}^{c+i\infty}\frac{\d\lambda}{2\pi i}\,e^{\lambda z}\,\hat{\F}(\lambda)$ for any $c>-1$; closing its contour to the left onto the cut, then rotating back to real time, recasts the mode function as a continuous superposition of plane waves weighted by the discontinuity of $\hat{\F}$ across the cut,
\begin{equation}
\F(z)=\frac{1}{2\pi}\int_1^\infty \d\lambda\;e^{-i\lambda z}\,\big[\mathrm{Disc}_{\lambda'}\hat{\F}(\lambda')\big]_{\lambda'=-\lambda}\,.
\label{integral-representation}
\end{equation}
The plane wave $e^{-i\lambda z}$ carries the entire time dependence; everything specific to the field, its mass and dynamics, sits in the discontinuity of $\hat{\F}$ across the cut, reflected by $\lambda\to-\lambda$ onto $\lambda\in(1,\infty)$ so that its endpoint $\lambda=1$ is the dual image of the early-time Bunch-Davies point at $\lambda=-1$.

What makes that weight computable is a second, equally elementary fact: a typical bulk equation of motion has coefficients polynomial in $z$, so under the transform it becomes a dual equation of motion in Laplace space, of the same order in $\lambda$. We make this explicit on the object of interest for a massive scalar in de Sitter, $\F(z)=H^{(1)}_{i\mu}(-z)/\sqrt{-z}$, the index $i\mu$ carrying the mass through $\mu=\sqrt{m^2/H^2-9/4}$: its de Sitter wave equation, a Bessel equation in $z$, maps to the Legendre equation
\begin{equation}
(\lambda^2-1)\,\hat{\F}''(\lambda)+2\lambda\,\hat{\F}'(\lambda)+\big(\mu^2+\tfrac14\big)\hat{\F}(\lambda)=0\,,
\end{equation}
of degree $i\mu-\tfrac12$: the time evolution of the massive field is thus traded for the Legendre equation in the Laplace variable $\lambda$. Its two solutions $P_{i\mu-1/2}$ and $Q_{i\mu-1/2}$ are singled out by their behaviour at the endpoints $\lambda=\pm1$, and Bunch-Davies selects the one regular at $\lambda=+1$. The singularity of the discarded solution there is the dual image of the negative-frequency modes $e^{+iz}$ absent from the adiabatic vacuum, an excited-state singularity excluded from the outset; the surviving solution is fixed in normalisation by the early-time data at the other endpoint $\lambda=-1$. The weight is then the Legendre function,
\begin{equation}
\frac{H^{(1)}_{i\mu}(-z)}{\sqrt{-z}}=-\sqrt{\tfrac{2}{\pi}}\,e^{\frac{\pi\mu}{2}+\frac{i\pi}{4}}\int_1^\infty\!\!\d\lambda\;e^{-i\lambda z}\,P_{i\mu-1/2}(\lambda)\,.
\label{Hankel-plane-waves}
\end{equation}
For a conformally coupled or a massless field the weight trivialises, $P_0=1$ and $P_1(\lambda)=\lambda$, and one recovers these elementary mode functions. Nothing in these steps is special to this field: any mode function whose equation of motion is polynomial in $z$ yields a dual equation of the same order, whose Bunch-Davies-regular solution provides the kernel entering in \eqref{integral-representation}.

\paragraph*{\bf Correlators in Laplace space.}
We work in de Sitter space, the leading description of the inflationary background, with conformally coupled external fields $\varphi$ coupled through polynomial interactions to massive scalars $\sigma$ exchanged internally. Conformally coupled legs are the natural starting point: the correlators of the primordial curvature perturbation follow from them by acting with weight-shifting operators and taking suitable soft limits of the external momenta~\cite{Jazayeri:2022kjy}. The equal-time correlators are evaluated at a late conformal time $\tau_0$ in the in-in (Schwinger-Keldysh) formalism~\cite{Weinberg_2005,Chen:2017ryl}, with the fields placed in the Bunch-Davies vacuum. The conformally coupled mode function is an exact plane wave up to a power factor, $\varphi(\tau;E)=iH\tau\,e^{-iE\tau}/\sqrt{2E}$, while a field of mass $m$ carries a Hankel mode function,
\begin{equation}
\sigma(\tau;s)=-i\frac{H\sqrt{\pi}}{2}\,e^{-\frac{\pi}{2}\mu+i\frac{\pi}{4}}(-\tau)^{3/2}H^{(1)}_{i\mu}(-s\tau)\,.
\end{equation}
A correlator is a sum of diagrams, each a product of nested conformal-time integrals over propagators built from these mode functions.

Inserting the plane-wave representation \eqref{Hankel-plane-waves} for every massive internal line of an in-in diagram trades each line for a plane wave. At a vertex $i$ the scale factor, the $C_i$ conformally coupled legs and the rescaling of the $M_i$ massive lines combine into a single power of conformal time, $\tau^{N_i}$ with $N_i=C_i-4+2M_i$, multiplying plane waves. The vertex time integral then reduces to a single flat-space integral of $\tau^{N_i}$ against plane waves: the conformally coupled legs supply the total external energy $E_i$, each adjacent massive line its rescaled energy $s_{ij}\lambda_{ij}$, and the Schwinger-Keldysh indices fix the signs and the $i\epsilon$ tilts. The power $\tau^{N_i}$ is reinstated by acting on the seed at $N_i=0$ with the differential operator $(-a_i i)^{N_i}\partial_{E_i}^{N_i}$; up to it the $\tau$ integral is elementary, of pure flat-space type, and the whole geometry, field content and dynamics survive in the $\lambda$ integrals against the Legendre kernels. The nested cosmological time integrals have thus collapsed onto their flat-space counterparts, and the whole computation proceeds with plane waves only, the kernels carrying all the departure from massless fields in Minkowski.

\paragraph*{\bf Diagrammatic rules.}
Performing the collapsed time integrals once and for all turns the construction into Laplace-space rules. Split a diagram into Schwinger-Keldysh contributions by assigning a branch index $a_i=\pm$ to each vertex, and orient its time-ordered internal lines. To every massive line joining vertices $i$ and $j$ attach a dual variable $\lambda_{ij}$ and the Legendre measure $\int_1^\infty\d\lambda_{ij}\,P_{i\mu_{ij}-1/2}(\lambda_{ij})$, the kernel of the exchanged field.

The flat-space time integrals then reduce to the energy denominators familiar from flat space. To each vertex assign the energy $V_i=E_i+\sum_j (\pm)\,s_{ij}\lambda_{ij}$, the external energy $E_i$ entering the vertex plus the rescaled internal energies $s_{ij}\lambda_{ij}$, the sign set by the line orientation (plus for outgoing or unordered, minus for ingoing). Integrating the nested plane waves yields one denominator per vertex,
\begin{equation}
\mathcal{V}_i=i\,a_i\Big(V_i+\!\!\sum_{j\,\text{earlier}}\!\!V_j\Big)+\epsilon\,,
\end{equation}
the inner sum running over vertices preceding $i$ along the ordering.

The de Sitter dressing is restored by two final operations: a differential operator $(-a_i i)^{N_i}\partial_{E_i}^{N_i}$ at each vertex reinstates the powers of the scale factor and of the conformally coupled legs, with $N_i$ given above, and one sums over branch assignments and time-orderings. The rescaled diagram of a given Schwinger-Keldysh contribution is then
\begin{equation}
\begin{aligned}
\tilde{\F}_{\{a_i\}}(\{E_i\},\{s_{ij}\})=\int_1^\infty\!\Big(\prod_{i\neq j}\d\lambda_{ij}\,P_{i\mu_{ij}-1/2}(\lambda_{ij})\Big)\\
\times\prod_i(-a_i i)^{N_i}\partial_{E_i}^{N_i}\,\frac{1}{\mathcal{V}_i}\,,
\end{aligned}
\end{equation}
a flat-space rational function of the energies integrated against one Legendre kernel per massive line; the correlator follows from the $\tilde{\F}_{\{a_i\}}$ by a kinematic rescaling given in~\cite{Belrhali:2026laplace}. Tree and loop diagrams are treated alike, loops adding only loop-momentum integrals that reduce to flat-space ones.

\paragraph*{\bf Massive single exchange.}
The method shows its full force on the paradigmatic massive single exchange, two conformally coupled legs meeting at each vertex and a scalar of mass parameter $\mu$ flowing through the internal line, with external energies $E_1,E_2$, internal energy $s$, and $\lambda_u=E_1/s$, $\lambda_v=E_2/s$. The rules of the previous section involve no scale-factor derivatives here ($N_1=N_2=0$) and deliver the correlator as the real part of two Schwinger-Keldysh components $\tilde{\F}_{++}$ and $\tilde{\F}_{+-}$, each a double integral of one Legendre kernel per massive line. The factorised component collapses through the dispersive integral $\int_1^\infty\d\lambda\,P_{i\mu-1/2}(\lambda)/(\lambda+z)=\frac{\pi}{\cosh(\pi\mu)}P_{i\mu-1/2}(z)$, each layer integrating to a Legendre function, so that $s^2\tilde{\F}_{+-}=\frac{\pi^2}{\cosh^2(\pi\mu)}P_{i\mu-1/2}(\lambda_u)P_{i\mu-1/2}(\lambda_v)$, while the time-ordered one keeps both line integrals, 
\begin{equation}\label{eq: double integration single exchange}
\begin{aligned}
s^2\,\tilde{\F}_{++}=-\!\int_1^\infty\!\!\!\int_1^\infty\!\!\d\lambda\,\d\tilde\lambda\;\frac{P_{i\mu-1/2}(\lambda)\,P_{i\mu-1/2}(\tilde\lambda)}{\lambda_T+\lambda-\tilde\lambda-i\epsilon}\\
\times\Big[\frac{1}{\lambda_u+\lambda-i\epsilon}+\frac{1}{\lambda_v+\lambda-i\epsilon}\Big]\,,
\end{aligned}
\end{equation}
with $\lambda_T=\lambda_u+\lambda_v$ the total energy and $\lambda,\tilde\lambda$ the dual variables carried by the exchanged line on the two branches. This is the form the diagrammatic rules produce for any diagram, and it has a twofold virtue.

First, this integral representation makes the singularity structure manifest, each energy singularity of the correlator arising as a flat-space pole of the rational integrand that reaches the boundary $\lambda=1$ of the Laplace integration domain, dual to the asymptotic past. At vanishing total energy, $\lambda_T\to0$, the total-energy denominator $\lambda_T+\lambda-\tilde\lambda$ is the only factor coupling the two layers: the principal value of the $\tilde\lambda$ integral develops a $\log(\lambda+\lambda_T-1)$, itself singular as the remaining variable reaches the boundary, so that the non-analyticity is generated at the corner where all the dual variables sit at $\lambda=1$, the Laplace-space image of every vertex being pushed to the early-time past. There the Legendre kernels trivialise, $P_{i\mu-1/2}(1)=1$, and, with $\lambda_v\to-\lambda_u$ in this limit, the leading behaviour is $s^2\,\Re\tilde{\F}_{++}\to\frac{2}{1-\lambda_u^2}\,\lambda_T\log\lambda_T$, a mass-independent coefficient proportional to the flat-space amplitude for the exchange of a massless field~\cite{Maldacena:2011nz,Raju:2012zr}. At a vanishing partial energy, $\lambda_u\to-1$, the roles are exchanged and a single integration variable is involved: it is now the rational factor $1/(\lambda_u+\lambda)$, an ordinary flat-space energy denominator, whose pole reaches the endpoint $\lambda=1$ and produces a $\log(1+\lambda_u)$, while the other layer, performed with the dispersive integral, leaves the Legendre kernel itself at a rotated argument, $P_{i\mu-1/2}(e^{i\pi}\lambda_v)$. This is the Laplace-space three-point function of two conformally coupled and one massive field, in agreement with the factorisation of correlators into a lower-point correlator and a lower-point amplitude near partial-energy singularities (see e.g.~\cite{Baumann:2020dch,Goodhew:2020hob,Jazayeri:2021fvk}). Although illustrated here on the single exchange, this reading holds diagram by diagram: the energy singularities of an arbitrary correlator are inherited from flat space through the very same boundary mechanism.

\begin{figure}[t]
\centering
\includegraphics[width=\columnwidth]{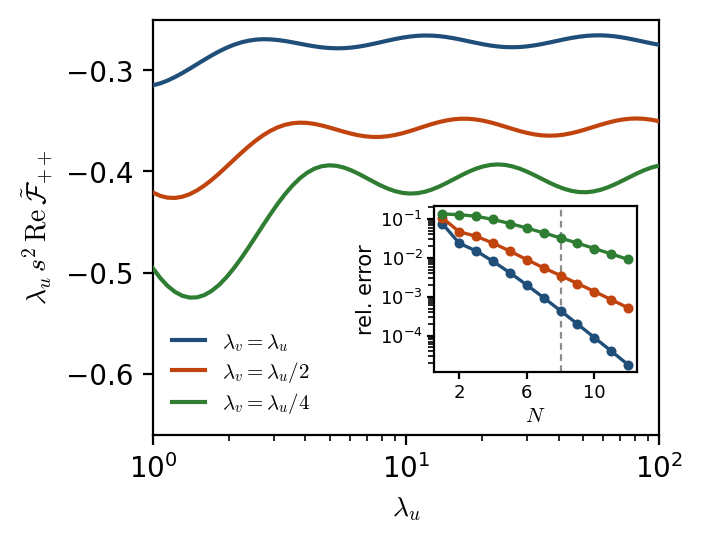}
\caption{The single-exchange correlator $s^2\,\Re\tilde{\F}_{++}$ for $\mu=2$, rescaled by $\lambda_u$ to expose the oscillatory cosmological collider signal on top of the smooth effective-field-theory background, along three slices of fixed ratio $\lambda_v/\lambda_u$. The curves are the master series truncated at total order $2n+m\le N=8$ ($25$ terms). Manifestly symmetric under $\lambda_u\leftrightarrow\lambda_v$, this single expansion covers the whole domain at once, including the equal-energy configuration $\lambda_v=\lambda_u$ that sits at the boundary between the orderings handled by separate expansions in other approaches. \emph{Inset}: the relative error of the truncation, maximised over each slice, falls geometrically with the order $N$ (dashed line: the value used in the main panel), fastest at equal energy and slowing towards the hierarchical configurations, but rapid throughout.}
\label{fig:convergence}
\end{figure}

Second, the same representation evaluates in closed form. Performing one of the two line integrals with the same dispersive integral brings $\Re\tilde{\F}_{++}$ to a single integral over a Legendre $Q$ kernel,
\begin{equation}
\begin{aligned}
s^2\,\Re\tilde{\F}_{++}=2\!\int_1^\infty\!\!\d\lambda\;P_{i\mu-1/2}(\lambda)\,\Re\,Q_{i\mu-1/2}(\lambda+\lambda_T)\\
\times\Big[\frac{1}{\lambda+\lambda_u}+\frac{1}{\lambda+\lambda_v}\Big]\,.
\end{aligned}
\end{equation}
Two expansions then reduce the integrand to elementary blocks: the large-argument expansion of the $Q$ kernel, $Q_{i\mu-1/2}(z)=\sqrt{\pi}\,\frac{\Gamma(1/2+i\mu)}{\Gamma(1+i\mu)}(2z)^{-1/2-i\mu}\sum_n a_n z^{-2n}$, and the expansion of the rational factors in the ratios $\lambda_{u,v}/(\lambda+\lambda_T)<1$.  Each resulting term is the generalised dispersive integral $T_\rho(\lambda_T)\equiv\int_1^\infty\d\lambda\,P_{i\mu-1/2}(\lambda)/(\lambda+\lambda_T)^\rho$, a one-parameter extension of the dispersive integral that evaluates in closed form to a Gauss hypergeometric function. The correlator then resums into a single closed-form series,
\begin{widetext}
\begin{equation}
\begin{aligned}
&s^2\,\Re\tilde{\F}_{++}= 2\sqrt{\pi}\;\Re\!\left[\frac{\Gamma(\tfrac12+i\mu)}{\Gamma(1+i\mu)}\,2^{-\frac12-i\mu}\sum_{n,m=0}^{\infty}a_n\left(\lambda_u^{\,m}+\lambda_v^{\,m}\right)T_{2n+m+\frac32+i\mu}(\lambda_T)\right]\,,\\[4pt]
&a_n=\frac{\left(\tfrac14+\tfrac{i\mu}{2}\right)_n\left(\tfrac34+\tfrac{i\mu}{2}\right)_n}{\left(1+i\mu\right)_n\,n!}\,,\qquad
T_\rho(\lambda_T)=\frac{2^{1-\rho}\,\Gamma\!\left(\rho-\tfrac12\pm i\mu\right)}{\Gamma(\rho)}\;{}_2\tilde F_1\!\left(^{\rho-\frac12+i\mu,\ \rho-\frac12-i\mu}_{\qquad\quad \rho};\tfrac{1-\lambda_T}{2}\right)\,.
\end{aligned}
\end{equation}
\end{widetext}
resummed in the ratios $\lambda_{u,v}/(1+\lambda_T)$ and manifestly symmetric under $\lambda_u\leftrightarrow\lambda_v$. A single expansion thus converges geometrically across the entire kinematic domain, with no patching of the two orderings handled by separate, asymmetric expansions in earlier approaches~\cite{Arkani-Hamed:2018kmz,Qin:2022fbv,Werth:2024mjg}: the equal-energy configuration $\lambda_u=\lambda_v$, which lies precisely at the boundary between those two regions, is here treated like any other and is in fact where convergence is fastest, the rate decreasing towards hierarchical configurations. Its relative error falls geometrically with the truncation order and is already negligible at the order $N=8$ ($25$ terms) shown in Fig.~\ref{fig:convergence}. This single representation delivers the full correlator at once, capturing on the same footing the effective-field-theory background and the oscillatory cosmological collider signal of the massive exchange~\cite{Chen:2009zp,Chen:2009we,Noumi:2012vr,Arkani-Hamed:2015bza}, well visible in Fig.~\ref{fig:convergence}.

\paragraph*{\bf Generality.}
The explicit setup above assumed polynomial interactions and the exchange of massive scalars, but the construction is tied to neither. The plane-wave skeleton is universal, and the only field-specific input is the Laplace transform of the corresponding mode function, the kernel that dresses the elementary plane waves into it; whenever that transform is known, the diagrammatic rules carry over unchanged and only the kernel is replaced. This makes the method apply, in increasing order of departure from the explicit setup, to essentially any theory of interest in primordial cosmology.

Wavefunction coefficients are treated on the same footing as correlators, with no Schwinger-Keldysh index and a bulk-to-bulk propagator whose extra component is handled identically, with the corresponding diagrammatic rules given in~\cite{Belrhali:2026laplace}. Derivative interactions change nothing essential: spatial gradients only add kinematic factors and powers of conformal time, and temporal derivatives act directly on the plane-wave seeds, the analysis again proceeding with plane waves only. Reduced sound speeds are absorbed, up to overall factors, by rescaling the external energies and internal momenta, since each Laplace representation goes through with $z=c_s k\tau$. Spinning fields are reached as well: the transverse modes of a spin-1 field with a chemical potential are Whittaker functions, which time-dependent couplings turn into twisted Whittaker functions, whose plane-wave representation is given in~\cite{Belrhali:2026laplace}. The same logic ultimately reaches beyond de Sitter altogether, in any background where the Laplace transform of the relevant bulk function can be computed, which can always be done numerically when no closed form is available.

\paragraph*{\bf Discussion.}
A Laplace transform has recast the curved-space evolution of each cosmological mode into plane waves dressed by a kernel, collapsing the nested in-in time integrals onto elementary flat-space ones. This leaves simple rules that produce the Laplace-space integrand directly from a diagram and, on the massive single exchange, a closed-form series convergent across the whole kinematic domain. The difficulty of the curved-space problem does not vanish, it is relocated. The entire departure from flat space is repackaged into one rigid, universal structure: known kernels, each fixed by a dual equation of motion, integrated over the fixed domain $\lambda\in[1,\infty)$ against flat-space rational denominators. That this structure is powerful and not merely elegant is what the single exchange proves, and it is the structure shared by essentially every correlator of primordial cosmology, whose computation is thereby brought to one common flat-space form.

This recasting is more than a computational convenience: it is genuinely ``from flat space'', and that carries conceptual weight. The clearest evidence is the singularity structure, where the total- and partial-energy singularities are inherited directly from their flat-space origins, the massless amplitude surfacing at the total-energy corner and the three-point function at a partial-energy edge. Yet these singularities are only one face of a broader fact. Where earlier flat-space constructions dressed scattering amplitudes and were confined to massless or conformally coupled fields, here flat space enters through massless correlators and reaches the massive exchanges in full generality. The Laplace structure thus provides a way to elucidate the intricate analytic structure of cosmological correlators by uplifting the comparably much simpler one of flat-space correlators.

Because only the kernel changes from one theory to the next, the same machinery is ready for the cosmological collider, spinning fields and departures from scale invariance alike. Furthermore, radiative corrections reduce to loop momentum integrals over flat-space quantities, simply
dressed by Laplace-space kernels. A single representation, uniform and rapidly convergent across kinematics, is moreover well suited to scanning shapes, theories and parameters. Should this efficiency persist for the more intricate correlators beyond the single exchange, it would hand observational cosmology a practical tool to confront theory with data. The reach may run further still: wherever a curved background becomes asymptotically flat, modes return to ordinary Minkowski plane waves, and a Laplace construction of the same kind should apply, around black holes for instance, with the radial coordinate playing the role of our conformal time. Built on so elementary a fact and yet this general, the Laplace approach opens a new avenue on the correlators of the early universe, at once a calculational engine and a way of understanding their structure.

\paragraph*{\bf Acknowledgements.} We thank Denis Werth for initial collaboration and Guillaume Faye, Sebastian Garcia-Saenz, Austin Joyce and Zhong-Zhi Xianyu for useful discussions. We are also grateful for the feedback of the many participants in several scientific events where Nathan Belrhali presented this work while it was in preparation: the program 
	\href{https://indico.ijclab.in2p3.fr/event/11373/}{CoBALt} held at the Institut Pascal at Universit\'e Paris-Saclay with the support of the program ``Investissements
	d'avenir'' ANR-11-IDEX-0003-01, the \href{https://indico.in2p3.fr/event/35602/}{TUG workshop} at IPhT Saclay, the 41st annual IAP symposium \href{https://indico.iap.fr/event/36/}{Inflation 2025}, the \href{https://earlyuniverse.discussingresearch.com/}{Early Universe from Home} 2026 online conference, and the \href{https://indico.cern.ch/event/1556583/overview}{PONT 2026} conference in Avignon.

\bibliography{references-Laplace-clean.bib}

\end{document}